\documentclass[%
 reprint,
superscriptaddress,
 amsmath,amssymb,
 aps,
prx,
]{revtex4-1}

\usepackage{graphicx}
\usepackage{dcolumn}
\usepackage{bm}
\usepackage[colorlinks=true, allcolors=blue]{hyperref}
\usepackage{siunitx}
    \DeclareSIUnit \counts{Cts}
    \DeclareSIUnit \electron{e}
    \DeclareSIUnit \eV{eV}
    \DeclareSIUnit \molar{M}
\usepackage{braket}
\usepackage{amsmath} 
\usepackage{amssymb}
\usepackage{amsfonts}

\usepackage[inline]{trackchanges}
\addeditor{DH}
\addeditor{LB}
\addeditor{RG}

\begin{document}

\preprint{APS/123-QED}

\title{Amplified Sensitivity of Nitrogen-Vacancy Spins in Nanodiamonds using \\ All-Optical Charge Readout}

\author{David A. Hopper}
\affiliation{Quantum Engineering Laboratory, Department of Electrical and Systems Engineering, University of Pennsylvania, Philadelphia Pennsylvania 19104, USA}
\affiliation{Department of Physics and Astronomy, University of Pennsylvania, Philadelphia Pennsylvania 19104, USA}
\author{Richard R. Grote}
\affiliation{Quantum Engineering Laboratory, Department of Electrical and Systems Engineering, University of Pennsylvania, Philadelphia Pennsylvania 19104, USA}
\author{Samuel M. Parks}
\affiliation{Quantum Engineering Laboratory, Department of Electrical and Systems Engineering, University of Pennsylvania, Philadelphia Pennsylvania 19104, USA}
\author{Lee C. Bassett}
\email{lbassett@seas.upenn.edu}
\affiliation{Quantum Engineering Laboratory, Department of Electrical and Systems Engineering, University of Pennsylvania, Philadelphia Pennsylvania 19104, USA}

\date{\today}

\begin{abstract}
 Nanodiamonds containing nitrogen-vacancy (NV) centers offer a versatile platform for sensing applications spanning from nanomagnetism to \textit{in-vivo} monitoring of cellular processes. In many cases, however, weak optical signals and poor contrast demand long acquisition times that prevent the measurement of environmental dynamics. Here, we demonstrate the ability to perform fast, high-contrast optical measurements of charge distributions in ensembles of NV centers in nanodiamonds and use the technique to improve the spin readout signal-to-noise ratio through spin-to-charge conversion.
A study of 38 nanodiamonds, each hosting 10-15 NV centers with an average diameter of 40 nm, uncovers complex, multiple-timescale dynamics due to radiative and non-radiative ionization and recombination processes.  Nonetheless, the nanodiamonds universally exhibit charge-dependent photoluminescence contrasts and the potential for enhanced spin readout using spin-to-charge conversion. We use the technique to speed up a $T_1$ relaxometry measurement by a factor of five.

\end{abstract}

\pacs{Valid PACS appear here}
\maketitle
\section{Introduction}
Nitrogen-vacancy (NV) centers embedded in nanodiamonds combine the advantages of an optically-addressable, room-temperature spin qubit \cite{Awschalom2013} with the nanoscale dimensions and flexible surface chemistry of diamond nanoparticles \cite{Mochalin2012,Schirhagl2014}.  Recent proof of concept demonstrations of their quantum sensing capabilities include magnetic \cite{LeSage2013} and thermal \cite{Kucsko2013} imaging in living cells; detection of electrochemical potentials \cite{Grotz2012,Petrakova2015,Karaveli2016}, paramagnetic molecules \cite{Steinert2013, Tetienne2013}, and pH levels \cite{Rendler2017} in solution; and investigations of ferromagnetism on the nanoscale \cite{Rondin2013,Pelliccione2016,Tetienne2016,Andrich2017}. All of these sensing modalities demand strong interactions between NV qubits and a target environment outside the diamond, for which nanoparticles are ideal.  However, high impurity levels and uncontrolled surface states in nanodiamonds \cite{Chang2008,Dantelle2010} degrade the NV's spin and charge stability compared to the situation in bulk diamond, leading to signal averaging issues and limited sensitivity for nanodiamonds due to poor optical contrast for charge \cite{Karaveli2016} and spin \cite{Bogdanov2017} readout by photoluminescence (PL) techniques. 

In conventional PL-based spin readout, an intense \SI{532}{\nano\meter} probe produces slightly more PL photons for the NV's $m_s=0$ ground-state spin sublevel as compared to $m_s=\pm1$ for the first $\approx$\SI{300}{\nano\second} of illumination.  However, recently established spin-to-charge conversion (SCC) protocols offer a more flexible approach \cite{Hopper2016,Shields2015}.  SCC utilizes the intersystem crossing dynamics of the NV$^-$ excited state to protect one spin state from an intense ionization pulse, leading to a spin-dependent charge distribution. A subsequent charge-selective optical pulse detects the resulting distribution, with the potential for dynamical tuning of power and duration for optimum efficiency \cite{Danjou2016}.

\begin{figure}[h]
\includegraphics[scale=1]{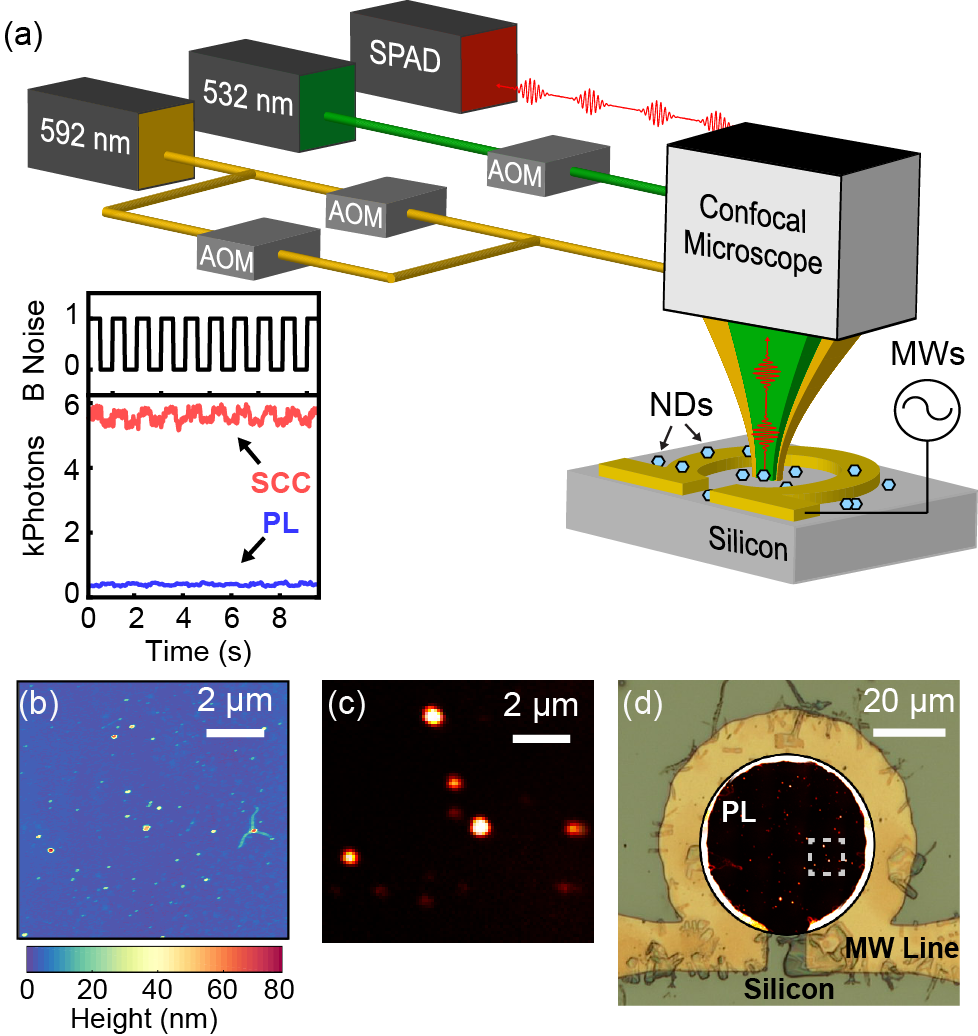}
\caption{\label{fig:intro}\textbf{Charge and Spin Control of Individual Nanodiamonds} (a) Experimental set-up consisting of a two-color home-built confocal microscope imaging nanodiamonds dispersed on a microwave antenna device fabricated on silicon (SPAD, single-photon avalanche diode; AOM, acousto-optic modulator; NDs, nanodiamonds; MWs, microwaves). Inset: Experimental demonstration of a $T_1$ spin relaxometry measurement using PL (blue) and SCC (red) depicting the SNR amplification for a nanodiamond agglomerate. Atomic force microscope measurements (b) and PL images (c) of the region boxed in (d) indicate the presence of individual nanodiamonds.}
\end{figure}
Here, we present all-optical protocols for high-contrast charge readout and SCC of NV ensembles in nanodiamonds as a means for boosting the signal-to-noise ratio (SNR) for charge and spin measurements as compared to conventional PL techniques, as seen in Fig.~\ref{fig:intro}(a). An investigation of the optically induced charge dynamics suggests that NVs in nanodiamonds milled from type Ib high-pressure, high-temperature (HPHT) diamond are prone to multiple non-radiative ionization (negative to neutral) and recombination (neutral to negative) pathways that are not observed in bulk, type IIa diamond.  We attribute these dynamics to tunneling transitions involving nearby impurity sites.  Despite these complications, we confirm on a sample of 38 individual nanodiamonds and several larger nanodiamond ensembles that high-contrast charge readout and SCC protocols, which to date have only been demonstrated in high-purity bulk diamond \cite{Hopper2016,Shields2015}, are still feasible and advantageous for quantum sensing protocols. 

For single NVs in bulk, type IIa diamond, strong charge-dependent optical contrasts facilitate high-fidelity, single-shot measurements of the NV's charge state \cite{Waldherr2011, Aslam2013,Hopper2016}. These measurements rely on the large energy difference in the zero phonon lines of the neutral charge state (NV$^0$, \SI{2.156}{\eV}) and  the negative charge state (NV$^-$, \SI{1.945}{\eV}) as well as the fact that the ionization and recombination mechanisms are two-photon processes \cite{Waldherr2011a}. Recently, charge readout has been extended to ensembles of NVs in type Ib bulk diamond \cite{Dhomkar2016,Jayakumar2016}, which is typically used to produce nanodiamonds with NV centers \cite{Chang2008,Dantelle2010}. Charge dynamics in bulk type Ib diamond are complicated by impurity-related charge transfer mechanisms \cite{Manson2005,Jayakumar2016}, and the situation in nanodiamonds is even less well understood. A few studies have aimed to maximize the NV$^-$ population under continuous illumination \cite{Rondin2010,Havlik2013,Berthel2015} or to measure charge-dependent stimuli using nanodiamonds \cite{Petrakova2015,Karaveli2016}. Improved charge readout techniques could vastly improve the sensitivity of such measurements. 



High-contrast charge readout is a prerequisite of SCC protocols for enhanced spin readout \cite{Shields2015,Hopper2016}.  Here we consider all-optical readout, although electrical charge measurements are also possible when the NVs are incorporated in a junction structure \cite{Hrubesch2017,Gulka2017,Brenneis2015}. For all-optical SCC, the charge readout produces larger SNR for longer readout times. Thus, applications with long measurement cycles, such as $T_1$ sensing schemes, stand the most to gain from spin SNR improvements. NV $T_1$ relaxometry has enabled gadolinium-based biological sensing \cite{Pelliccione2014,Schirhagl2014,Rendler2017}, direct imaging of nanoscale magnetism \cite{Schmid-Lorch2015, Tetienne2016, Pelliccione2016}, as well as microwave-free nanoscale electron spin resonance \cite{Hall2016} due to the ground state spin's sensitivity to fast fluctuating magnetic fields \cite{Steinert2013,Tetienne2013}. Since the $T_1$ times of NVs in nanodiamonds can take values ranging from \SI{3}{\micro\second} to \SI{4}{\milli\second} \cite{Tetienne2013}, measurement acquisition times can vary over three orders of magnitude depending on the NV under study.
A striking example of this signal averaging bottleneck is the recent demonstration of scanning $T_1$ relaxometry imaging \cite{Tetienne2016}, which demands \SI{1}{\second} dwell time per \SI{25}{\nano\meter}-wide pixel, resulting in \SI{2.5}{\micro\meter} wide scans taking 150 minutes to acquire. 

\section{Results}
Fluorescent nanodiamonds milled from HPHT Ib diamond (Ad\'{a}mas Nanotechnologies) were drop cast onto silicon substrates patterned with titanium gold wires for microwave control [Fig.~\ref{fig:intro}(a)]. The concentration of the nanodiamond solution was chosen to limit particle aggregation such that isolated nanodiamonds could be resolved in a confocal microscope using \SI{532}{\nano\meter} (\SI{2.33}{\eV}) excitation and PL collection. The presence of single and few nanodiamonds was confirmed by comparing atomic force microscope (AFM) scans [Fig.~\ref{fig:intro}(b)] with 2D confocal PL scans [Fig.~\ref{fig:intro}(c)]. The AFM scans in Fig.~\ref{fig:intro}(b) exhibit a height distribution spanning $39\pm\SI{19}{\nano\meter}$, in agreement with the vendor's specification. According to the vendor, each nanodiamond contains 10-15 NVs, although the variation in PL brightness across nanodiamonds suggests a broader distribution \cite{Supplemental}. In addition to the \SI{532}{\nano\meter} pump laser, a continuous-wave \SI{592}{\nano\meter} (\SI{2.09}{\eV}) laser is split into two arms for independent power and timing control and subsequently recombined with the excitation path [Fig.~\ref{fig:intro}(a)] for use as a pump or probe for charge state control and measurement. Collected PL was spectrally filtered between \SI{650}{\nano\meter} and \SI{775}{\nano\meter} to suppress emission originating from the NV$^0$ charge state and directed to a single-photon avalanche diode. The collected PL signal, $S(\tau)$, defined as the time-dependent photon detection rate as a function of the probe duration, $\tau$, is proportional to the population of NV centers in the negative charge state. Further details on the sample preparation and measurement setup can be found in the supplemental information \cite{Supplemental}.

\begin{figure*}[t]
\includegraphics[scale=1]{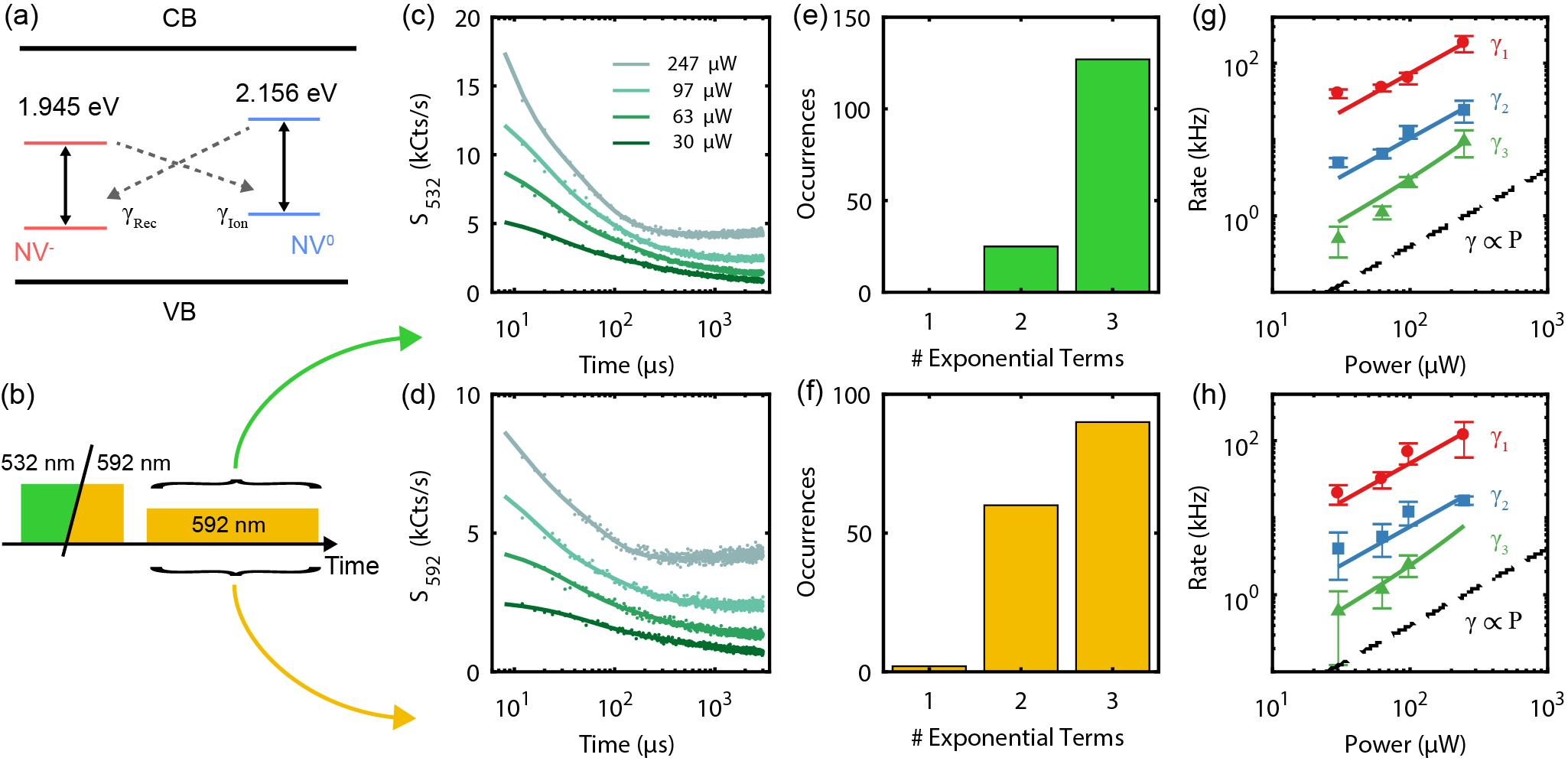}
\caption{\textbf{Optical Charge Dynamics} (a) Electronic level diagram depicting the NV$^-$ and NV$^0$ zero phonon line energies along with charge conversion processes (dotted arrows). (CB, conduction band; VB, valence band) (b) Experimental timing diagram. Panels (c)-(h) show examples of the time-correlated PL response (c,d; points) along with tri-exponential fits (c,d; lines), distributions of the optimal number of exponential terms (e,f), and the probe-power dependence of the rates $\gamma_1$, $\gamma_2$, and $\gamma_3$ of the tri -exponential fits in c and d where the top (g) and bottom (h) panel corresponds to initialization with \SI{532}{\nano\meter} or \SI{592}{\nano\meter} illumination, respectively. The dashed line in (g, h) signifies the expected shape of a linear power scaling.\label{Fig:extrinsic_dynamics}}
\end{figure*}

To study the optically induced charge dynamics of the NV ensembles in nanodiamonds, we preferentially populate either the NV$^-$ or NV$^0$ charge states using \SI{532}{\nano\meter} or \SI{592}{\nano\meter} pump beams, respectively \cite{Aslam2013}, and read out the resulting NV$^-$ population with a low-power \SI{592}{\nano\meter} probe beam. Figure~\ref{Fig:extrinsic_dynamics} summarizes the results of these measurements, in which different initial conditions and probe powers serve to map out the dynamical response due to different ionization and recombination mechanisms [Fig.~\ref{Fig:extrinsic_dynamics}(a)]. The timing sequence is depicted in Fig.~\ref{Fig:extrinsic_dynamics}(b), and the time-correlated PL response due to four different \SI{592}{\nano\meter} probe intensities following pumping by either \SI{532}{\nano\meter} and \SI{592}{\nano\meter} light are shown in Figs.~\ref{Fig:extrinsic_dynamics}(c) and \ref{Fig:extrinsic_dynamics}(d), respectively. These representative data exhibit multi-timescale and occasional non-monotonic behavior that is observed to varying degrees across all 38 nanodiamonds in this study \cite{Supplemental}. 

We fit all of the data to an empirical multi-exponential function of the form:
\begin{equation}
S_{\lambda}(\tau) = C^{(\lambda)}_0+\sum_{k=1}^{n}C_k^{(\lambda)}\exp[-\gamma_k \tau],
\label{Eqn:exp_fit_func}
\end{equation}
where $\lambda\in \{532,592\}$ signifies initialization by green or orange pump beams and $n$ is the number of exponential terms. The Akaike Information Criterion is used to determine the value of $n$ required to best represent the observed data \cite{Spiess2010, Supplemental}. We find that all of the measurements for 38 nanodiamonds can be fit as either single ($n=1$), bi- ($n=2$), or tri-exponential ($n=3$) functions with the coefficient labels ordered such that $\gamma_1 > \gamma_2 > \gamma_3$.  The solid lines in Figs.~\ref{Fig:extrinsic_dynamics}(c, d) are examples of fits using a tri-exponential model.
The distribution of optimized exponential number ($n$) for a total of 152 time-correlated probe responses for both \SI{532}{\nano\meter} and \SI{592}{\nano\meter} pump conditions are shown in Figs.~\ref{Fig:extrinsic_dynamics}(e, f), respectively. In a majority of cases, the tri-exponential model most accurately recreates the data. The relative increase of bi-exponential cases with a \SI{592}{\nano\meter} pump is presumably due to a larger portion of the NVs already close to the steady state following initialization with the same wavelength, which simplifies the dynamics. Nevertheless, the fact that dynamical behavior is observed at all following \SI{592}{\nano\meter} initialization is indicative of power-dependent ionization and recombination processes and charge relaxation in the dark, likely due to the lower thermodynamic stability of NV$^0$ compared to NV$^-$ \cite{Gaebel2006}.  The empirical multi-exponential model accounts for these multiple competing processes and for the fact that each nanodiamond contains an ensemble of NVs with different local environments due to the proximity of surfaces and other impurity states \cite{Jayakumar2016}.

A closer look at the fit results provides insight into the ionization and recombination mechanisms of the NV ensembles. Figures~\ref{Fig:extrinsic_dynamics}(g) and \ref{Fig:extrinsic_dynamics}(h) display the best-fit rates as a function of laser power for the data in Figs.~\ref{Fig:extrinsic_dynamics}(c) and \ref{Fig:extrinsic_dynamics}(d), respectively. Since the \SI{592}{\nano\meter} probe intensities are maintained below 6\% of the saturation power (\SI{4.3}{\milli\watt}), we expect the rates to exhibit a polynomial power dependence whose order depends on the number of photons involved in each ionization or recombination process \cite{Waldherr2011a}. In contrast to the case for single NVs in bulk, type IIa diamond, where ionization and recombination requires at least two photons with a wavelength of \SI{592}{\nano\meter} \cite{Waldherr2011a,Aslam2013,Hopper2016}, we observe a non-negligible linear component in the power scaling for all rates and initial conditions. The linear term points to the existence of a single-photon ionization or recombination mechanism. Similar behavior has been observed for NV ensembles in bulk, type-Ib diamond \cite{Manson2005}, where it is believed to result from tunneling of an electron or hole from the NV excited state to a nearby substitutional nitrogen trap \cite{Zvyagin2012}. By computing the excitation rate from a saturation curve, we estimate that $\approx$3\% of all cycling events result in a non-radiative tunneling event.

\begin{figure}[t]
\includegraphics[scale=1]{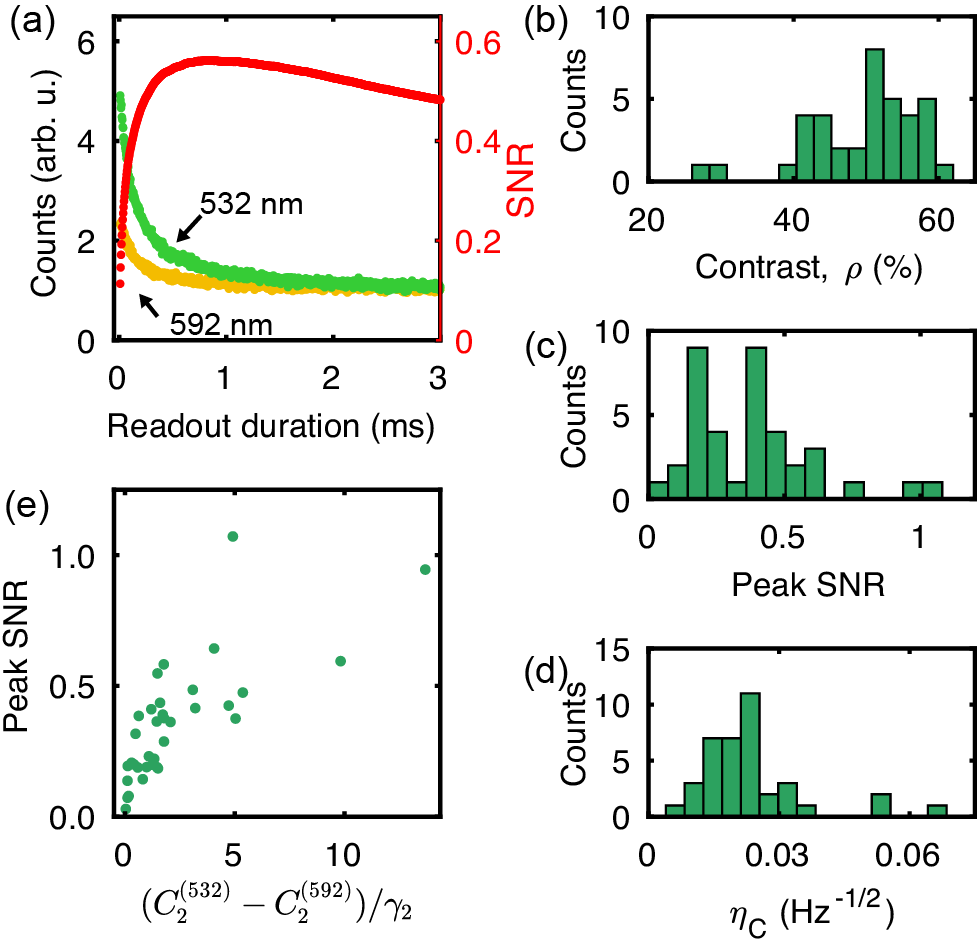}
\caption{\textbf{High-Contrast Optical Charge Readout} (a) Time-correlated PL traces for \SI{532}{\nano\meter} and \SI{592}{\nano\meter} initial conditions together with the single-shot SNR as a function of readout time. Panels (b)-(d) show distributions of the maximum contrast (b), peak SNR (c), and time-averaged SNR (d) for the 38 nanodiamonds studied. (e) Peak SNR as a function of the charge-dynamics figure of merit discussed in the text. \label{Fig:charge_readout}}
\end{figure}

Despite the complicated dynamics, pumping with \SI{532}{\nano\meter} or \SI{592}{\nano\meter} illumination still produces large differences in the charge populations that can be read out optically.  Fig.~\ref{Fig:charge_readout}(a) depicts how the different charge distributions manifest as a signal contrast within the time-correlated PL of a representative nanodiamond. Along with the photon counting data, we plot the corresponding single-shot charge measurement SNR as a function of readout duration, which is defined by
\begin{align}
\textrm{SNR}(\tau) = \frac{\alpha_{532}(\tau) - \alpha_{592}(\tau)}{\sqrt{\alpha_{532}(\tau) + \alpha_{592}(\tau)}},
\label{Eqn:SNR}
\end{align}
where $\alpha_\lambda(\tau)=\int_0^\tau S_\lambda(\tau')d\tau'$ is the total number of photons detected after probe duration $\tau$ following initialization with wavelength $\lambda$. Here we assume the noise is dominated by photon shot noise.  The SNR initially increases with $t$ as more photons are detected but eventually reaches a maximum before decreasing as the contrast vanishes and shot noise takes over.

To investigate the universality of this charge readout mechanism, in Figs.~\ref{Fig:charge_readout}(b-d) we plot the statistical distributions of various performance metrics calculated from the set of measurements on 38 nanodiamonds summarized in Fig.~\ref{Fig:extrinsic_dynamics}. For each nanodiamond, we calculate the initial optical contrast at the lowest probe power, 
\begin{align}
\rho = \left(1 - \frac{S_{592}(0)}{S_{532}(0)}\right)\times 100\%,
\label{Eqn:Contrast}
\end{align}
which reflects the difference in initial charge population.
The distribution of $\rho$, seen in Fig.~\ref{Fig:charge_readout}(b), exhibits a narrow peak around the mean contrast of $50\pm 7\%$. Notably, all of the observed values are lower than the ideal contrast of $\rho_\mathrm{bulk}=84\%$ expected for NVs in bulk, type-IIa diamond \cite{Aslam2013}.  We attribute this difference to the finite duration of our measurements and the more complicated local environment of NVs in nanodiamonds.  Nonetheless, every nanodiamond we studied exhibits a strong optical charge contrast. Figure~\ref{Fig:charge_readout}(c) shows the distribution of peak single-shot SNR values, optimized for readout power and duration. Here we find a much wider distribution with a mean SNR~$=0.38\pm0.22$. This large spread of values is not surprising given the widely varying nanodiamond brightness due to different NV ensemble sizes, together with variations in the charge dynamics during readout due to different local environments. Finally, for each nanodiamond we also calculate the time-averaged charge readout sensitivity, 
\begin{equation}
\eta_\mathrm{C} = \frac{\sqrt{\tau_R}}{\mathrm{SNR}(\tau_R)}.\label{eq:charge_sensitivity}
\end{equation}
Here we assume that the readout time, $\tau_R$, and the probe power are optimized to provide the maximum single-shot SNR. The charge sensitivity has units of $\mathrm{Hz}^{-1/2}$, and, assuming shot noise dominates the measurement uncertainty, dividing $\eta_\mathrm{C}$ by the square root of the total integration time, $T$, yields the minimum resolvable signal variation, $\delta = \eta_\mathrm{C}/\sqrt{T}$. The distribution of charge sensitivities is displayed in Fig.~\ref{Fig:charge_readout}(d). Twenty of the nanodiamonds surveyed exhibit $\eta<0.02$, meaning that we can resolve 2\% signal variations after one second of integration. Remarkably, despite the wide qualitative variation of optically induced charge dynamics, all nanodiamonds observed in this study showed contrasts between $26\%$ and $61\%$ and charge sensitivity better than 0.07~$\mathrm{Hz}^{-1/2}$. 

The qualitative variations of charge dynamics and distribution of charge readout performance metrics are not independent of each other. For example, a better charge measurement intuitively requires both a larger contrast to increase the signal amplitude and slower decay rates to allow for more detected photons. To test this hypothesis, we searched for correlations between metrics such as peak SNR, contrast, and sensitivity and particular parameters of our empirical models \cite{Supplemental}. Interestingly, the parameters most predictive of performance are the amplitude difference, $C_2^{(532)}-C_2^{(592)}$, and rate, $\gamma_2$, of the second exponential term in Eqn.~\ref{Eqn:exp_fit_func}. Figure \ref{Fig:charge_readout}(e) displays the strong correlation between the peak SNR and a combined figure of merit, $(C_2^{(532)}-C_2^{(592)})/\gamma_2$.  This analysis confirms our physical intuition and also offers an effective means of screening nanodiamonds for optimal performance as charge sensors.

\begin{figure*}[ht]
\includegraphics[scale=1]{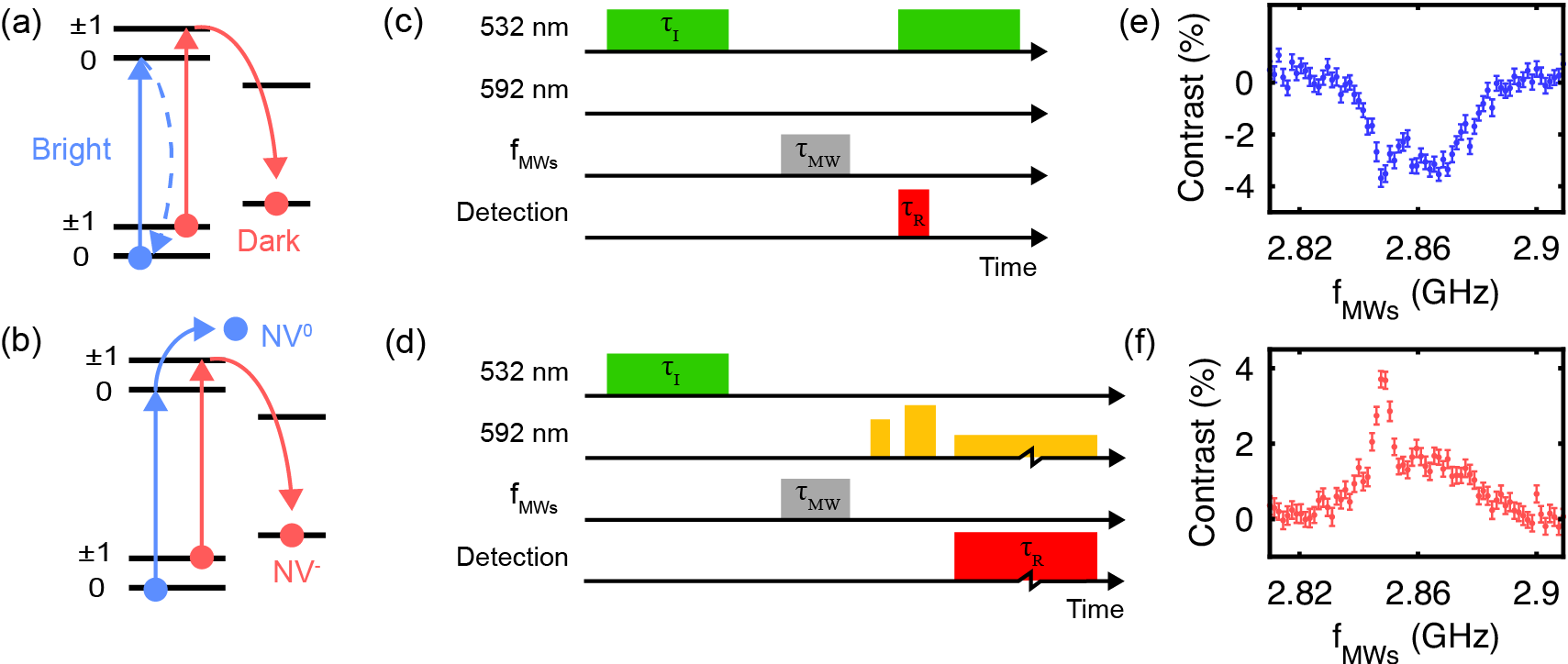}
\caption{\textbf{Spin-to-Charge Conversion in Nanodiamonds} (a,b) Diagrams of PL spin readout and the SCC process, respectively for NV$^-$. Solid lines represent pumped transitions, dotted lines represent radiative transitions. (c,d) Measurement timing diagrams for performing pulsed ESR using PL or SCC readout, respectively.  $f_{MWs}$ corresponds to the variable frequency of the microwave pulse, and the break in the readout pulse of (d) indicates a large span of time. $\tau_{\textrm{I}}$: initialization time, $\tau_{\textrm{MW}}$: microwave pulse time, $\tau_{\textrm{R}}$: readout duration. (e,f) Measured ESR spectra for a single nanodiamond using PL and SCC protocols, respectively.\label{Fig:ESR_comparison}}
\end{figure*}

The availability of a high-contrast charge measurement for nanodiamonds is crucial to achieve performance advantages using SCC readout protocols.  Figures \ref{Fig:ESR_comparison}(a) and \ref{Fig:ESR_comparison}(b) compare the mechanisms for spin readout using traditional PL and SCC, respectively. PL readout results from optically cycling the triplet manifold of NV$^-$, typically using a \SI{532}{\nano\meter} pump, which causes the $m_s=0$ spin projection to produce more photons (bright state) as compared to the $m_s=\pm1$ projection which is shelved \textit{via} the intersystem crossing into the metastable singlet (dark state). The essence of SCC is a timed optical pulse sequence that transfers the initial spin populations into either the triplet manifold (for $m_s=0$) or the singlet manifold (for $m_s=\pm1$) and then quickly ionizes the population selectively from one manifold or the other \cite{Shields2015,Hopper2016}.  Following this SCC procedure, a low-intensity, charge-selective probe pulse (\SI{592}{\nano\meter} in this work) detects the NV$^-$ population. Thus, the optical charge readout signal is correlated to the NV's initial spin state.

To verify the SCC mechanism in nanodiamonds, we performed pulsed electron spin resonance (ESR) measurements on a nanodiamond at zero magnetic field. The measurement timing diagrams for PL and SCC readout techniques are sketched in Figs.~\ref{Fig:ESR_comparison}(c) and \ref{Fig:ESR_comparison}(d), respectively. Both measurement sequences begin with a \SI{5}{\micro\second}, \SI{532}{\nano\meter} pump pulse to initialize the ensemble primarily into NV$^-$ and $m_s=0$. A weak, variable-frequency microwave pulse with a duration exceeding the inhomogeneous dephasing time ($\tau_{\textrm{MW}}=\SI{200}{\nano\second}$) is then applied to probe the ground state spin transition. In the case of PL readout, a second \SI{532}{\nano\meter} pulse is applied and photons are detected for the first \SI{300}{\nano\second}. For SCC, two pulses of \SI{30}{\milli\watt} \SI{592}{\nano\meter} light (a \SI{15}{\nano\second} shelving pulse followed \SI{25}{\nano\second} later by a \SI{50}{\nano\second} ionization pulse) \cite{Supplemental} are applied to perform the conversion process outlined in Fig.~\ref{Fig:ESR_comparison}(b). Due to the finite rise time of the AOM used to generate these pulses, the power of the shelving pulse is lower than that of the ionization pulse.  The same SCC pulse parameters were used for all nanodiamonds in this work. Finally, a \SI{430}{\micro\watt}, \SI{40}{\micro\second} probe pulse is applied with photon detection during the entire duration. Both SCC pulses are derived from one arm of the \SI{592}{\nano\meter} laser path shown in Fig.~\ref{fig:intro}(a) whereas the lower-power probe pulse is generated in the second arm.  The results are presented in Figs.~\ref{Fig:ESR_comparison}(e) and \ref{Fig:ESR_comparison}(f) for PL and SCC readout, respectively. Both spin measurement techniques show the typical response characterizing an NV ensemble with strong inhomogeneous broadening, confirming that SCC does indeed measure the spin state. The qualitative difference in curve shapes suggests that different NVs within the nanodiamond exhibit variations in their charge readout and SCC responses.

In order to quantify the potential improvement offered by SCC, we studied its spin readout performance in comparison to traditional PL. We calibrated the optimal measurement parameters for PL readout, and found that, in contrast to the situation in bulk diamond where optical excitation close to saturation is preferred, the optimal \SI{532}{\nano\meter} readout pulse was tuned to a factor of 4 below the saturation power for a duration of \SI{300}{\nano\second} \cite{Supplemental}. This observation agrees with other recent measurements of reduced spin SNR for NVs in nanodiamonds on sapphire substrates \cite{Bogdanov2017}. The non-NV PL contributes background levels $<1\%$ of the signal at the optimum spin readout power, so this cannot explain the anomalous SNR decrease. We believe the more complicated ionization and recombination mechanisms are the primary cause of this SNR decrease, since the probabilities of non-radiative charge transitions from the NV$^-$ excited state are comparable to those for the intersystem crossing \cite{Goldman2015a}.

\begin{figure}[t]
\includegraphics[scale=1]{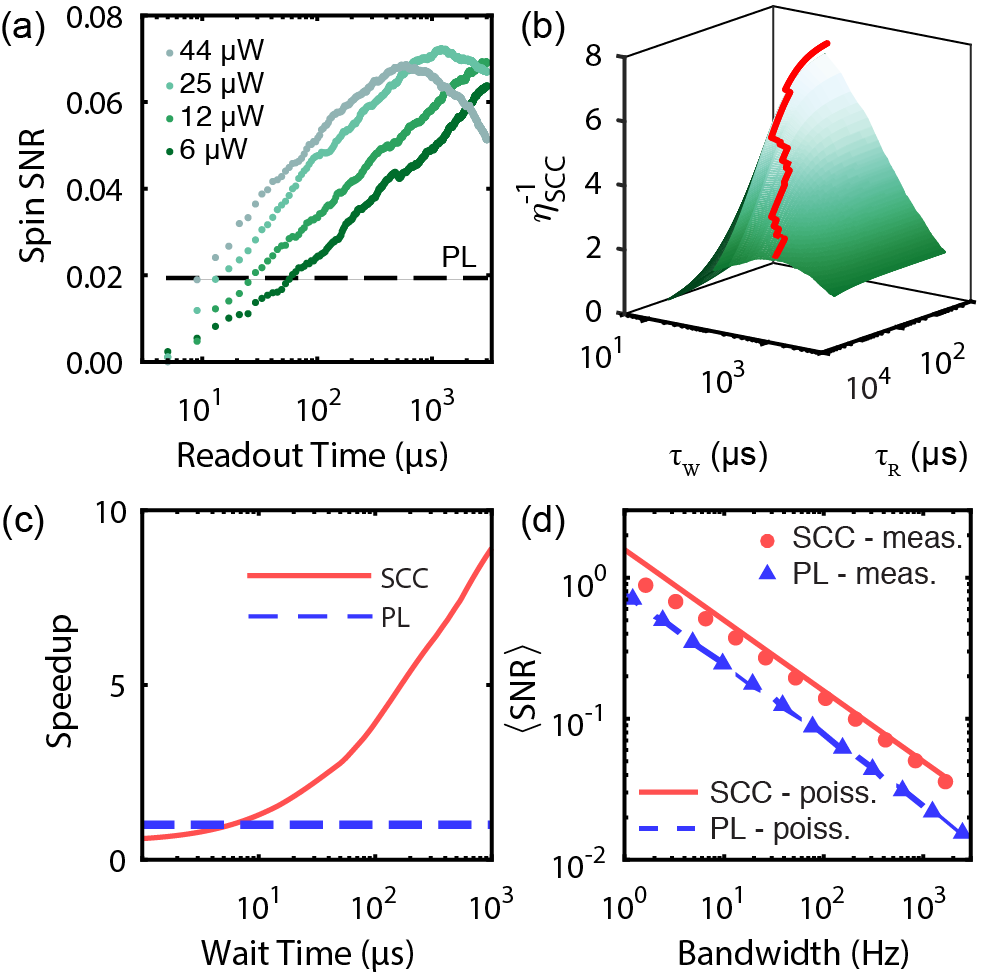}
\caption{\textbf{Performance of the SCC protocol} (a) Calibration of the SCC spin SNR (points). The measured PL SNR = 0.19 is indicated by a dashed line. (b) Inverse of the SCC sensitivity figure of merit (see text) as a function of wait time and readout duration, calculated from the data in (a). The red solid line indicates the optimum value, while the surface depicts how the sensitivity varies across all parameters. (c) Predicted speedup factor as a function of measurement wait time. (d) $T_1$ relaxometry measurement comparing the time-averaged SNR as a function of bandwidth for PL (blue triangles) and SCC (red circles) readout. The solid (dashed) lines indicate the predicted SNR calculated in a shot noise approximation for SCC (PL). Error bars are smaller than the data markers. \label{Fig:scc_demonstration}}
\end{figure}

Figure~\ref{Fig:scc_demonstration}(a) shows the resulting single-shot spin SNR for a single nanodiamond as a function of the probe pulse intensity ($P_{\textrm{R}}$) and duration ($\tau_{\textrm{R}}$). This particular nanodiamond has favorable charge and spin properties, with a peak charge-detection SNR~$=0.36$ and spin $T_1 = 780\pm$\SI{230}{\micro\second} \cite{Supplemental}. We observe that SCC out-performs PL readout whenever $\tau_\mathrm{R}>\SI{10}{\micro\second}$, with a factor of 3.8 improvement in SNR for $\tau_\mathrm{R}\approx\SI{1}{\milli\second}$.  As in the case of charge readout, for time-averaged measurements this presents an optimization tradeoff between the single-shot SNR and measurement duration.  Therefore, in analogy to eqn.~(\ref{eq:charge_sensitivity}), we calculate the time-averaged spin-readout sensitivity, 
\begin{equation}
\eta_\mathrm{SCC}(\tau_\mathrm{W}) = \max_{\tau_\mathrm{R},P_\mathrm{R}}\left(\frac{\sqrt{\tau_\mathrm{I} + \tau_\mathrm{W} + \tau_\mathrm{R}}}{\mathrm{SNR}(\tau_\mathrm{R},P_\mathrm{R})}\right),
\label{Eqn:time_averaged_SNR}
\end{equation}
where $\mathrm{SNR}(\tau_\mathrm{R},P_\mathrm{R})$ is the single-shot spin SNR at a given readout duration, $\tau_\mathrm{R}$, and power, $P_\mathrm{R}$.  Here we must include the total duration of the measurement sequence, composed of the constant initialization time, $\tau_\mathrm{I}$, and the variable waiting time (or, more generally, the spin-operation time), $\tau_\mathrm{W}$, in addition to $\tau_\mathrm{R}$.  Fig.~\ref{Fig:scc_demonstration}(b) depicts how the quantity $\left(\frac{\sqrt{\tau_\mathrm{I} + \tau_\mathrm{W} + \tau_\mathrm{R}}}{\mathrm{SNR}(\tau_\mathrm{R},P_\mathrm{R})}\right)^{-1}$ varies over the two dimensional measurement parameter space consisting of $\tau_\mathrm{W}$ and $\tau_\mathrm{R}$, once $P_\mathrm{R}$ has already been optimized. The red line tracing the ridge of the surface provides a visual indicator of the measured optimized experimental settings. 

Using this analysis, we can make a direct comparison between the performance of SCC and traditional PL protocols.  The PL readout sensitivity, $\eta_\mathrm{PL}(\tau_\mathrm{W})$, is calculated in a similar manner to eqn.~(\ref{Eqn:time_averaged_SNR}), except we assume the readout power and duration remain fixed at their single-shot optimal values. Then we calculate the speedup factor, \textit{i.e.}, the ratio of acquisition times required to achieve a common time-averaged SNR,
\begin{equation}
F(\tau_\mathrm{W}) = \left(\frac{\eta_\mathrm{PL}(\tau_\mathrm{W})}{\eta_\mathrm{SCC}(\tau_\mathrm{W})}\right)^2.
\end{equation}
The results of this analysis for the nanodiamond investigated are plotted in Fig.~\ref{Fig:scc_demonstration}(c). The break-even wait time, when $F=1$, occurs when $\tau_\mathrm{W}=\SI{6}{\micro\second}$, and by $\tau_\mathrm{W}=\SI{100}{\micro\second}$ the speedup has reached a factor of 5. Physically, this increase stems from the \si{\kilo\hertz} dynamics governing charge readout, which allows for long acquisition times and thus a large number of detected photons in each shot. In principle, $F$ will increase with $\tau_\mathrm{W}$ to a saturated value determined by the squared ratio of the single-shot SNRs ($\lim_{\tau_\mathrm{W}\rightarrow\infty}(F)\approx 14$ in this case). In practice, however, the range of useful values for $\tau_\mathrm{W}$ is limited by the sensing protocol of interest and ultimately by the spin lifetime.

To demonstrate performance improvements in a practical setting, we performed $T_1$ spin relaxometry measurements in which a fixed wait time of $\tau_\mathrm{W}=T_1/2=$\SI{390}{\micro\second} was used to sense magnetic disturbances in the local environment \cite{Tetienne2013}. The target signal consisted of low-power microwaves driven through the lithographic wire at \SI{2.87}{\giga\hertz} to simulate the presence of fast fluctuating magnetic fields around the nanodiamond. The presence of the microwaves reduces $T_1$ by more than an order of magnitude \cite{Supplemental}. Using the optimized SCC settings of $P_\mathrm{R}=\SI{44}{\micro\watt}$ and $\tau_\mathrm{R} = \SI{173}{\micro\second}$, we performed differential relaxometry measurements, comparing the signal with the microwaves on and off, for a range of total measurement bandwidths, $(1/T)$, where $T$ is the total measurement time. 
At each bandwidth, the time-averaged signal-to-noise ratio, $\langle\mathrm{SNR}\rangle=\Delta N/\sigma_N$, was calculated from the mean differential photon-counting signal, $\Delta N$, and the corresponding standard deviation, $\sigma_N$, for each technique.  This procedure was repeated 11 times to obtain statistics on the measured $\langle\mathrm{SNR}\rangle$. The results are plotted in Fig.~\ref{Fig:scc_demonstration}(d) along with the predicted variation of $\langle\mathrm{SNR}\rangle$ assuming only Poissonian noise contributions.  We observe an improvement from the SCC protocol by a factor $2.26\pm0.14$ corresponding to a speedup of $F = 5.11\pm0.63$, nearly independent of bandwidth.  Interestingly, while the measurements agree closely with the shot-noise prediction for the PL protocol,
the model including only Poissonian noise overestimates $\langle\mathrm{SNR}\rangle$ for the SCC protocol by 8$\%$.  We attribute this slight difference to additional noise introduced by the binomial nature of the SCC process \cite{Shields2015,Hopper2016}, although the effect is less prominent here compared to the case of single NVs due to inherent averaging over the ensemble of probed NVs. The slight decrease of $\langle$SNR$\rangle_{\textrm{SCC}}$ for bandwidths $\lesssim$\SI{10}{\hertz} is due to additional set up noise associated with the \SI{592}{\nano\meter} laser. 

Similar measurements to those previously described on large nanodiamond agglomerates containing several hundred NVs attest to the universality of the SCC protocol.  The supplementary information \cite{Supplemental} includes SCC calibration curves like Fig.~\ref{Fig:scc_demonstration} for such agglomerates, and Fig.~\ref{fig:intro}(a) depicts the amplification effect corresponding to a factor of 2.2 SNR improvement (factor of 5 speedup) for an agglomerate containing $\approx$100 NVs and for $\tau_\mathrm{W}=\SI{75}{\micro\second}$.

\section{Discussion}
The techniques described in this paper can be used to improve various schemes for nanoscale sensing using NVs in nanodiamonds.  The optical charge readout technique can readily be applied to measuring variations in the electrochemical potential surrounding nanodiamonds, produced for example by using an electro-chemical cell \cite{Karaveli2016} or functional groups on the nanodiamond surface \cite{Petrakova2015}. One potential future application of electrochemical sensing is the detection of neuron action potentials, which have $\approx$\SI{100}{\milli\volt} amplitudes and millisecond durations.  Comparing the time-averaged SNR measured by Karaveli \textit{et al.} \cite{Karaveli2016} using \SI{532}{\nano\meter} PL to our high-contrast charge measurements, we predict a factor of 5 improvement in charge sensitivity, corresponding to a factor of 25 speedup and the ability to detect milliVolt-scale variations in electrochemical potential on millisecond timescales. This can offer a microwave-free alternative to emerging techniques for action-potential sensing using NV ensembles in bulk diamond \cite{Barry2016} and the potential to extend these imaging modalities to \textit{in-vivo} studies. Future investigations of nanodiamond charge dynamics could employ more sophisticated optical pulse sequences, in which either the dark charge dynamics or the changes in steady state populations are measured quickly with large contrast to gain additional readout enhancements.

The improvements in spin readout using SCC offer a means to further improve nanodiamond magnetic sensing protocols, particularly for $T_1$ relaxometers where the spin-evolution time is long. 
For example, the factor of 5 speedup for $T_1$ sensing exhibited in Fig.~\ref{Fig:scc_demonstration}(d) would reduce the total acquisition time of the 2D relaxometry images demonstrated by Tetienne \textit{et al.} \cite{Tetienne2016} from 150 minutes to 30 minutes. These throughput improvements allow for the ability to measure more samples and also reduce the experimental complexity required to keep the imaging optics and sample stationary for such long periods of time. These results also motivate the investigation of other diamond NV platforms, such as bulk ensembles, shallow implanted NVs, and NVs coupled to waveguides or other photonic structures, which could achieve spin readout enhancements through the use of the time-averaged SCC protocol presented here. Recent results involving the coupling of NVs to nearby nuclear spins in nanodiamonds \cite{Knowles2016,Knowles2016a} also suggest that longer spin operation times will be required, which will further motivate the adoption of SCC to other NV-nanodiamond measurements. The promising improvement of $T_2$ times for shallow NVs \cite{FavarodeOliveira2017} suggests that these platforms will invariably encounter signal averaging issues as well, at which point SCC can offer major improvements.

\section{Conclusion}
We have developed all-optical protocols to amplify the charge and spin readout signals of NV ensembles in nanodiamonds for quantum sensing applications. A preliminary study of the optically induced charge dynamics suggests that the local environment of each NV within a given nanodiamond modulates the dynamics. Additional ionization and recombination mechanism that are not present for single NVs in high-purity diamond are consistent with the idea of tunneling between the NV excited state and nearby charge traps, although the intriguing dark dynamics warrants further investigation. A sampling of 38 nanodiamonds demonstrated the universality of high-contrast charge readout for these particles.  We further demonstrated a simplified two-color SCC protocol for nanodiamonds that provides spin readout enhancements in the context of a $T_1$ relaxometry measurement, resulting in a factor of 5 reduction in measurement acquisition time. These results provide a straightforward method for improving state-of-the-art quantum sensors beyond the limits already achieved using conventional PL spin readout. Furthermore, the improved sensing of electrochemical potentials motivates the development of nanodiamond charge sensors for measuring action potentials and local chemical potentials \textit{in-vivo}.
\bibliography{nd_scc_bibliography.bib}
\end{document}


\preprint{APS/123-QED}

\title{Supplemental Information for ``Amplified Sensitivity of Nitrogen-Vacancy Spins in Nanodiamonds using All-Optical Charge Readout''}
\author{David A. Hopper}
\affiliation{Quantum Engineering Laboratory, Department of Electrical and Systems Engineering, University of Pennsylvania, Philadelphia Pennsylvania 19104, USA}
\affiliation{Department of Physics and Astronomy, University of Pennsylvania, Philadelphia Pennsylvania 19104, USA}
\author{Richard R. Grote}
\affiliation{Quantum Engineering Laboratory, Department of Electrical and Systems Engineering, University of Pennsylvania, Philadelphia Pennsylvania 19104, USA}
\author{Samuel M. Parks}
\affiliation{Quantum Engineering Laboratory, Department of Electrical and Systems Engineering, University of Pennsylvania, Philadelphia Pennsylvania 19104, USA}
\author{Lee C. Bassett}
\affiliation{Quantum Engineering Laboratory, Department of Electrical and Systems Engineering, University of Pennsylvania, Philadelphia Pennsylvania 19104, USA}
\email{lbassett@seas.upenn.edu}

\date{\today}

\maketitle


\section{Sample Preparation}
Nanodiamonds were purchased from Adamas Nanotechnologies, item No. ND-15NV-40nm, and were reported to have a mean diameter of \SI{40}{\nano\meter} with approximately 10-15 NVs per ND. To produce single and few nanodiamonds through drop casting, the ND slurry of concentration \SI{1}{\milli\gram\per\milli\liter} was diluted by 4 orders of magnitude followed by horn sonication to break up agglomerates. The diluted solution was immediately drop cast onto O$_2$-plasma-cleaned silicon substrates and allowed to dry in atmosphere. The presence of isolated single and few nanodiamonds was confirmed by correlating atomic force microscope (AFM) measurements (MFP-3D atomic-force scanning probe, Asylum) and photolu\textrm{min}escence (PL) maps, as can be seen in the main text. For the studies involving agglomerates of nanodiamonds, the original concentration (\SI{1}{\milli\gram\per\milli\liter}) was dropcast directly onto a cleaned silicon substrate. The estimated number of NVs in the agglomerate was calculated by comparing saturation count rates of single nanodiamonds and the agglomerates.

\begin{figure}[b]
\includegraphics[scale=1]{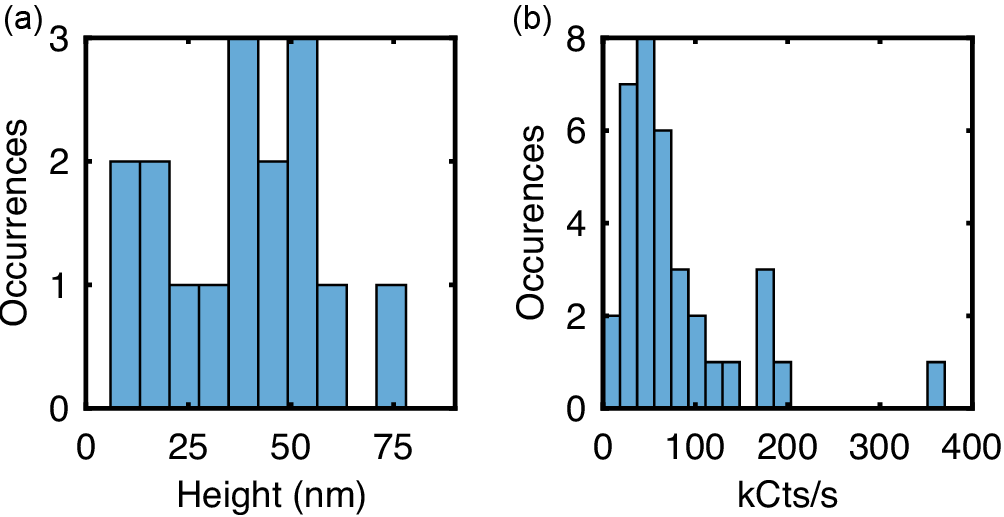}
\vspace{-10pt}
\caption{\label{sfig:bright_size}\textbf{Nanodiamond Variability} (a) Distribution of height of particles from the AFM scan in the main text showing single particle level heights. (b) Distribution of ND brightnesses at a fixed illu\textrm{min}ation power (above saturation) measured in this study.}
\end{figure}

The nanodiamonds exhibited variability in both size and consequently brightness, corresponding to the number of NVs contained in each nanodiamond. We show the size distribution of the AFM scan from the main text (Fig.1(b)), as well as the overall brightness distribution for all 38 nanodiamonds in SFigure~\ref{sfig:bright_size}. The mean height is around \SI{40}{\nano\meter} as expected. The variation in sizes of the nanodiamonds can have an effect on the charge dynamics due to the differing NV-to-surface distances.

\section{Confocal Microscope Details}
Nanodiamonds are imaged with a home-built confocal microscope with two excitation sources. Continuous-wave \SI{532}{\nano\meter} (Gem 532, Laser Quantum) and \SI{592}{\nano\meter} (VFL-592, MPB Communications, Inc.) lasers are gated with acousto-optic modulators (AOMs, 1250c, Isomet) with rise times of \SI{30}{\nano\second}. The \SI{592}{\nano\meter} laser is split with a beamsplitter (BS025, Thorlabs) into two arms to provide independent power control (NDC-50C-4, Thorlabs and 5215, Newport) and temporal gating. The beams are recombined with another beamsplitter (BS028, Thorlabs). The \SI{532}{\nano\meter} beam line is double passed through the AOM, which improves the extinction ratio to $>$\SI{60}{\decibel} and eliminates unwanted exposure to low power light which will cycle the defects' charge state between NV$^-$ and NV$^0$. The \SI{532}{\nano\meter} and \SI{592}{\nano\meter} beams are combined with a dichroic filter, co-aligned on a fast steering mirror (FSM, OIM101, Optics in Motion) and imaged through a 4$f$ lens configuration onto the back of an objective (Olympus MPlanFL N 100x, 0.9 NA). The collected photoluminescence is filtered to select for NV$^-$ in the 650-\SI{775}{\nano\meter} band, and focused onto a \SI{50}{\micro\meter}-diameter-core multi-mode optical fiber (M42L01, Thorlabs) that is connected to a single photon avalanche diode (Count-20c-FC, Laser Components). Photon detection events are recorded with a data acquisition card (DAQ-6323, National Instruments), which also functions as the global experimental clock. An arbitrary waveform generator (AWG520, Tektronix) controls the microwave pulse ti\textrm{min}g, optical pulse ti\textrm{min}g, and photon count gating. Microwaves are either supplied by lithographically patterned titanium-gold wires, or a gold bond wire laid across the silicon substrate. A signal generator (SG384, Stanford Research Systems) connected to a \SI{16}{\watt} amplifier (ZHL-16W-43-S+, \textrm{min}iCircuits) sources the microwaves for ground state spin control.

\section{Charge Dynamics Variability}
\begin{figure}
\includegraphics[scale=1]{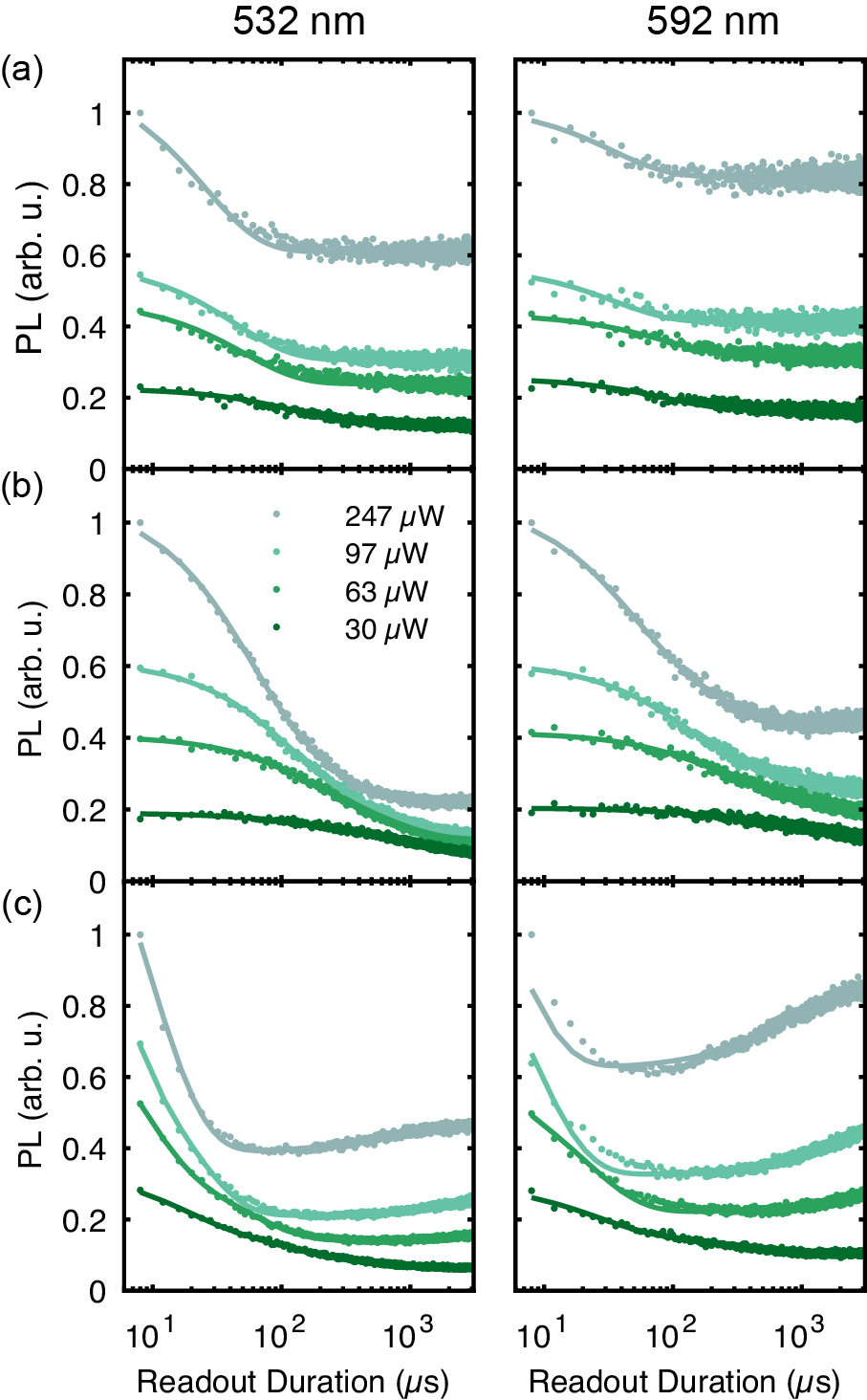}
\caption{\textbf{Optically Induced Charge Dynamics Variability} Time correlated PL responses for three nanodiamonds with \SI{532}{\nano\meter} and \SI{592}{\nano\meter} initial conditions for three nanodaimonds that exhibit different optimal models of (a) a single-exponential, (b) a bi-exponential, and (c) a tri-exponential. Data points are dots, solid lines are the fits.\label{sfig:pl_curves}}
\end{figure}

To supplement the example PL decay curves presented in the main text, we include the curves from three other nanodiamonds that show qualitatively different behavior (SFig.~\ref{sfig:pl_curves}). Both initialization conditions as well as the four different illu\textrm{min}ation powers are shown. Three different nanodiamonds are shown, each of which provides an example for the three different models($n=1,2$, or 3) considered in this study. The non-monotonic behavior is especially pronounced in SFig.~\ref{sfig:pl_curves}(c) for both initial conditions. Of particular note is the \SI{592}{\nano\meter}-pump response which shows that at higher powers the tri-exponential model fails to fully capture the data at short readout durations. This suggests that we would need to invoke higher numbers of exponentials to describe this PL response. However, this case was only present for a few nanodiamonds studied at the highest probe powers.

\section{AIC Model Selection}
We select the most approriate model for a given PL decay curve by calculating the Akaike Information Criterion (AIC). The AIC does not provide absolute goodness of fit, but rather ranks the models in accurately explaining the data \cite{Spiess2010}. The AIC can be calculated using the expression
\begin{align}
\textrm{AIC} = 2p - 2\textrm{ln}(L),
\end{align}
Where $p$ is the number of parameters in the model and $L$ is the likelihood function of the model. For the case of a nonlinear fit with normally distributed errors, the maximum log likelihood is given by
\begin{align}
\textrm{ln}(L) = 0.5\left(-N\left(\textrm{ln}(2\pi) + 1 - \textrm{ln}(N) + \textrm{ln}\left(\sum_{i=1}^{n}x_i^2\right)\right)\right),
\end{align}
where $N$ is the total number of data points and $x_i$ are the residuals \cite{Spiess2010}. Once the AIC value is calculated for each model, the model with the lowest AIC value is the one that best explains the observed data. Due to the information theory foundations upon which the AIC is formulated, we can also calculate how likely the other models are to actually be the correct model using the weight of evidence, defined as
\begin{align}
w_i = \frac{\exp(-(\textrm{AIC}_{\textrm{min}} - \textrm{AIC}_i)/2)}{\sum_{i=1}^k\exp(-(\textrm{AIC}_{\textrm{min}} - \textrm{AIC}_i)/2)},
\end{align}
where AIC$_{\textrm{min}}$ is the minimum AIC value deter\textrm{min}ed from the list, AIC$_i$ is the AIC value for the $i$ model, and $k$ is the total number of models. To lessen the effect of over-parameterization, we set a threshold of $w_i\leq0.05$. Thus if a lower complexity model has at least a 5$\%$ chance of accurately describing the measured data, we opt for the simpler model.

\begin{figure}[b]
\includegraphics[scale=1]{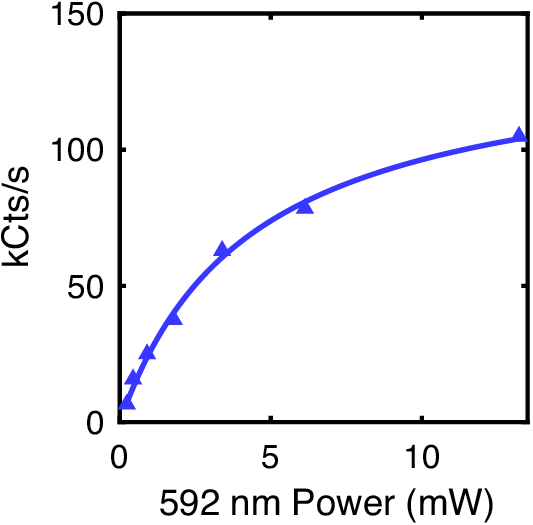}
\caption{\label{sfig:orange_sat_curve} \textbf{592~nm Saturation Curve}}
\end{figure}

\begin{figure*}[]
\includegraphics[scale=1]{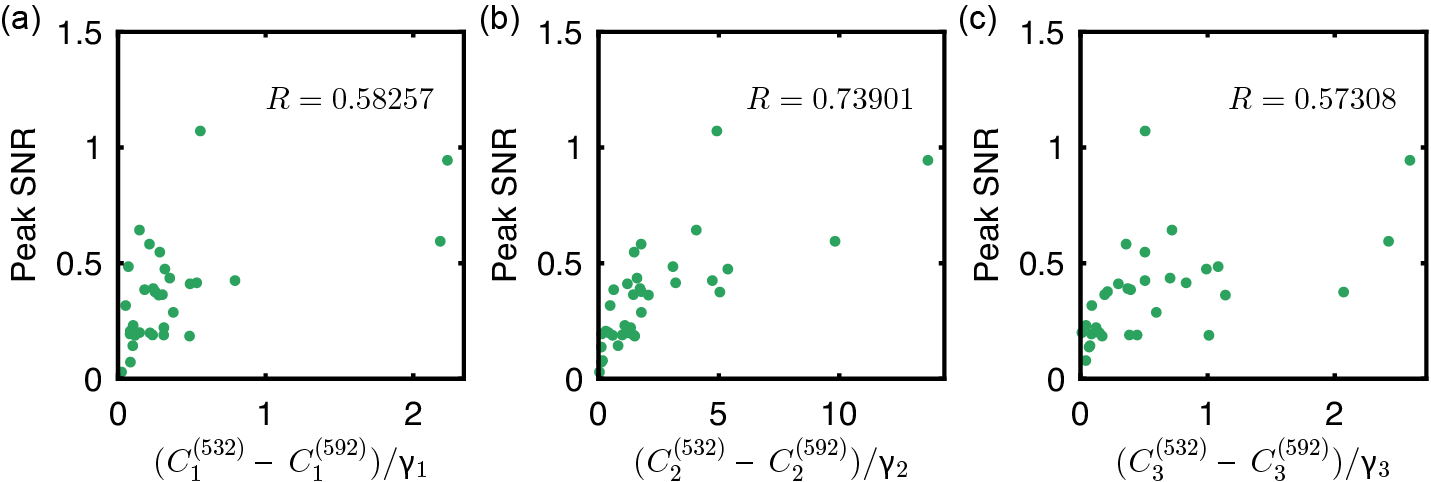}
\vspace{-15pt}
\caption{\textbf{Charge Readout SNR Correlations} Presented are a few correlations using the best fit values of our multi-exponential model. The first (a), second (b) and third (c) coefficient and rates are plotted against the peak SNR. The correlation coefficients are displayed in each plot.\label{sfig:correlation}}
\end{figure*}

\section{592~nm Saturation Curve}
To confirm that our \SI{592}{\nano\meter} probe pulse is in the weak excitation regime, we perform a saturation curve by changing the probe power and measuring the steady state PL rate (SFig.~\ref{sfig:orange_sat_curve}). This also allows us to back out the excitation rate at a given power, which provides a measure for comparing the excitation and charge decay rates as stated in the main text assuming that at saturation the excitation rate is equal to the spontaneous decay rate (1/\SI{20}{\nano\second}$^{-1}$). All nanodiamonds studied exhibited similar saturation behavior. The data in SFig.~\ref{sfig:orange_sat_curve} is fit to the expression
\begin{align}
\textrm{PL} = \textrm{PL}_{\textrm{sat}}\frac{1}{1 + \frac{I_{\textrm{sat}}}{I}},\label{eqn:sat_curve}
\end{align}
where $\textrm{PL}_{sat}$ is the saturation count rate, $I_{sat}$ is the saturation power, and $I$ is the illumination power. Our best fit values are; $C_{sat}=138\pm$\SI{14}{\kilo\counts\per\second} and $I_{sat}=4.3\pm$\SI{1}{\milli\watt} where error bars correspond to $90\%$ confidence intervals. This count rate is somewhat lower than expected for 10-15 NVs using a high-NA objective, but we attribute this to a combination of spectral filtering and the nanodiamonds' proximity to an absorptive high-index substrate (silicon, $n=3.8$).  Importantly, the collection efficiency does not affect our primary results for amplified spin readout via SCC, since increasing the count rate would improve both spin readout techniques in a similar way.

\section{Charge-readout SNR Correlations}
As stated in the main text, we searched for correlations between the empirical parameters (measurable quantities and best fit values to dynamical models) to best predict the peak charge-state SNR of a nanodiamond. We found that a good predictor of SNR was a figure of merit defined as
\begin{align}
\textrm{FOM}_i = (C_i^{532} - C_i^{592})/\gamma_i,
\end{align}
where $i$ signifies which exponential term the best-fit values come from. The relationship between these quantities for all three exponential terms, for the lowest illumination power, are presented in SFig.~\ref{sfig:correlation}, along with the Pearson correlation coefficient. It can be seen that the second exponential term acts as the best predictor for the peak SNR with a correlation coefficient of $R=0.73901$. While it seems that the third term ($i=3$) should best predict the SNR due to this term having the slowest rate, our results show otherwise. One reason for this could be that, although the third exponential is required to explain the dynamics, the magnitude of the contrast induced by this is significantly smaller than the second term. Other potential predictors, such as brightness or nanodiamond size, were not as predictive as the second exponential figure of merit. This result suggests that it is the charge dynamics that govern the performance of the readout, as opposed to shear number of NVs in the nanodiamond.

\section{SCC Parameter Calibration}
\begin{figure*}
\includegraphics[scale = 1]{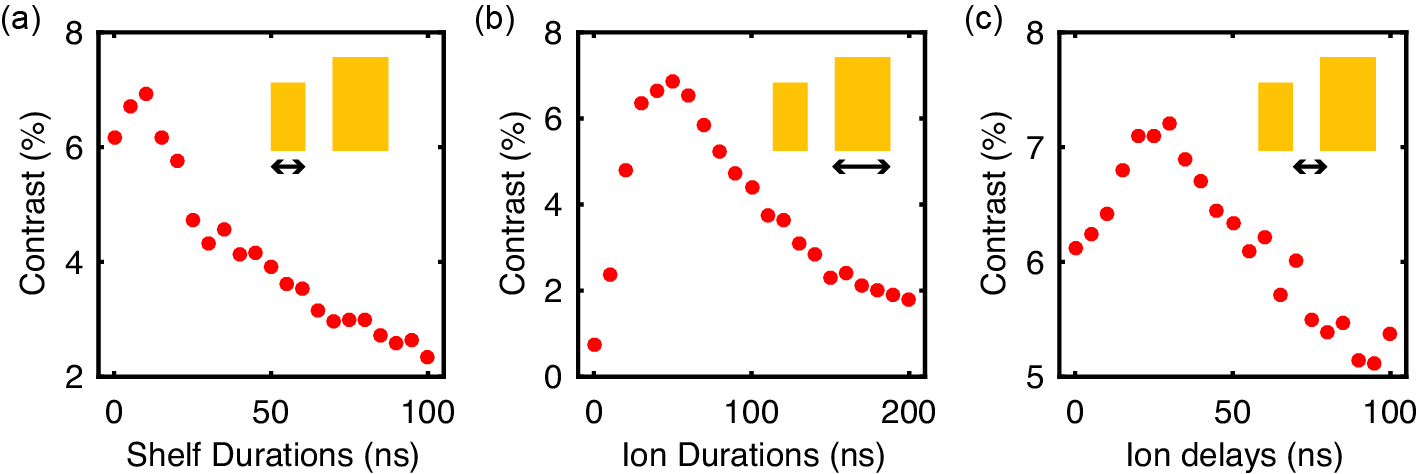}
\caption{\textbf{SCC Pulse Parameters} SCC contrast is maximized for (a) shelf duration, (b) ionization duration, and (c) ionization delay. Pulse diagrams inset depict the corresponding ti\textrm{min}g. Error bars are smaller than the markers.\label{sfig:scc_params}}
\end{figure*}

We performed a parameter sweep over the three relevant SCC pulse parameters of shelve-pulse duration, ionization-pulse duration, and ionization delay (SFig.~\ref{sfig:scc_params}) to find the optimal settings for maximum contrast. The optimum times found were a shelve-pulse duration of \SI{15}{\nano\second}, an ionization-pulse duration of \SI{50}{\nano\second}, and an ionization delay of \SI{25}{\nano\second}. We used a \SI{30}{\milli\watt} \SI{592}{\nano\meter} laser for both the shelving and ionization pulse. Due to the finite turn on time of the AOM, the shelving pulse is effectively a lower power than the ionization pulse. These parameters were optimized on an agglomerate of nanodiamonds, and we found that they produced satisfactory spin contrasts across all nanodiamonds studied, suggesting that the intersystem crossing rates, excited state lifetimes, and the fast photo-ionization are NV independent. Due to the optical power of the ionization pulse exceeding the saturation power of the ensemble, the dominant photo-ionization mechanism is the sequential absorption of two photons to eject and electron to the conduction band, as opposed to the tunneling mechanism involving nearby impurities discussed in the main text.

\section{PL Spin-Readout SNR Calibration}
\begin{figure}[h]
\includegraphics[scale = 1]{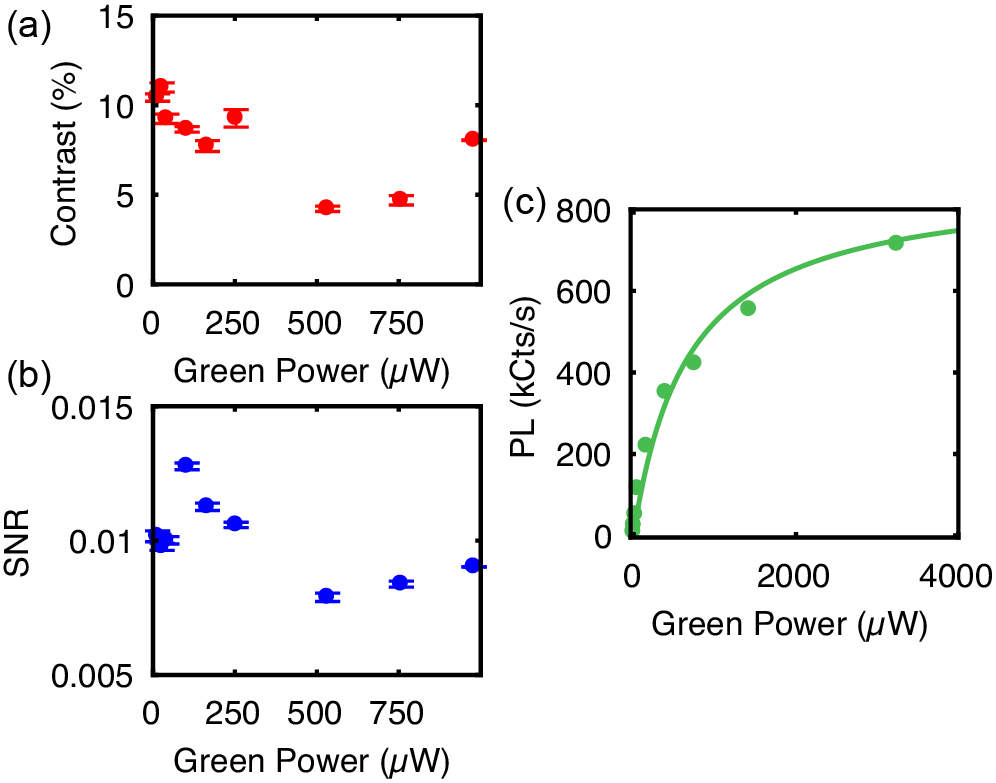}
\caption{\textbf{Reduced PL Spin SNR} (a) Spin contrast as a function of green readout power, optimized over readout duration. (b) Spin SNR as a function of green power, optimized over readout duration. (c) Saturation curve, depicting the saturation power of $670\pm240$~\si{\micro\watt}. \label{fig:pl_snr}}
\end{figure}

To ensure that we make a valid comparison between SCC and traditional PL spin readout, we varied the parameters for PL spin readout, particularly the readout duration and \SI{532}{\nano\meter} power. The results of this calibration, along with a saturation curve for reference, are presented in SFig.~\ref{fig:pl_snr}. We note that a global optimum in the PL-readout SNR occurs for powers about 6-fold below the saturation power ($I_{\textrm{sat}}=670\pm\SI{240}{\micro\watt}$, $PL_{\textrm{sat}}=870\pm\SI{100}{\kilo\counts}$) but with a typical readout duration of \SI{300}{\nano\second}. Similar decreases in spin contrast and SNR for nanodiamonds with increasing \SI{532}{\nano\meter} excitation power has recently been reported \cite{Bogdanov2017}.

\section{Single ND $T_1$ Time}
\begin{figure}[h]
\includegraphics[scale = 1]{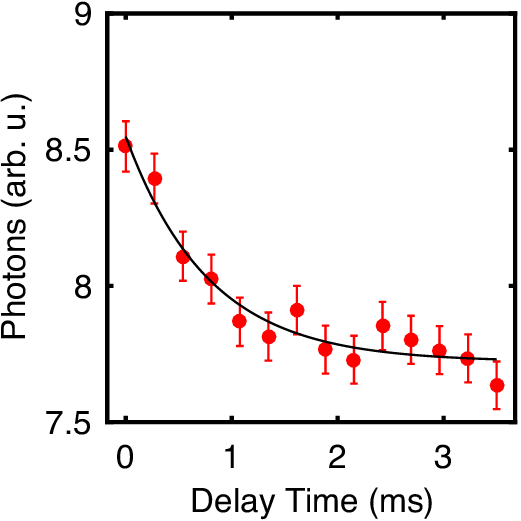}
\caption{\textbf{Single ND $\bm{T_1}$} $T_1$ measurement for the nanodiamond mentioned in the text for which the SCC calibration was performed. Solid black line is the fit to the data.\label{sfig:single_nd_t1}}
\end{figure}

The $T_1$ measurement of the nanodiamond used in the SCC demonstration in the main text is displayed in SFig.~\ref{sfig:single_nd_t1}. The data is fit to a single exponential curve and we find a longitudinal spin lifetime of $T_1=780\pm$\SI{230}{\micro\second}.

\section{$T_1$ Modulation}
\begin{figure}
\includegraphics[scale = 1]{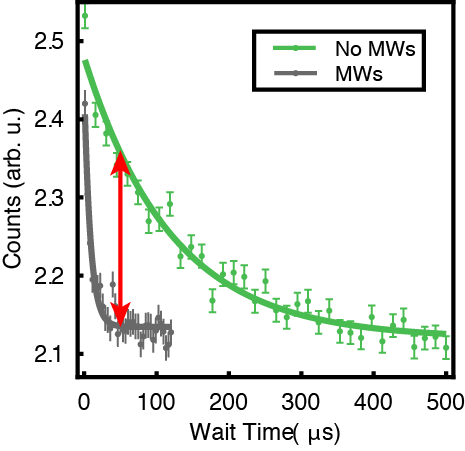}
\caption{\textbf{$\bm{T_1}$ Modulation} $T_1$ measurement with and without low power (20 dBm) microwaves. The red arrow signifies the optimal contrast that is obtained when using a wait time of $T_1/2$.\label{sfig:t1_mod}}
\end{figure}

Due to the NV center's sensitivity to fast fluctuating fields near the ground state splitting, we can simulate an environmental signal by applying low power microwaves at \SI{2.87}{\giga\hertz} during a $T_1$ measurement delay time. As can be seen in SFig.~\ref{sfig:t1_mod}, the characteristic time is reduced from \SI{150}{\micro\second} to below \SI{10}{\micro\second} with the microwaves applied for an agglomerate of nanodiamonds used for the qualitative SNR comparison presented in the main text (Fig.~1(a)).

\section{Agglomerate SCC Calibration}
\begin{figure} [h]
\includegraphics[scale=1]{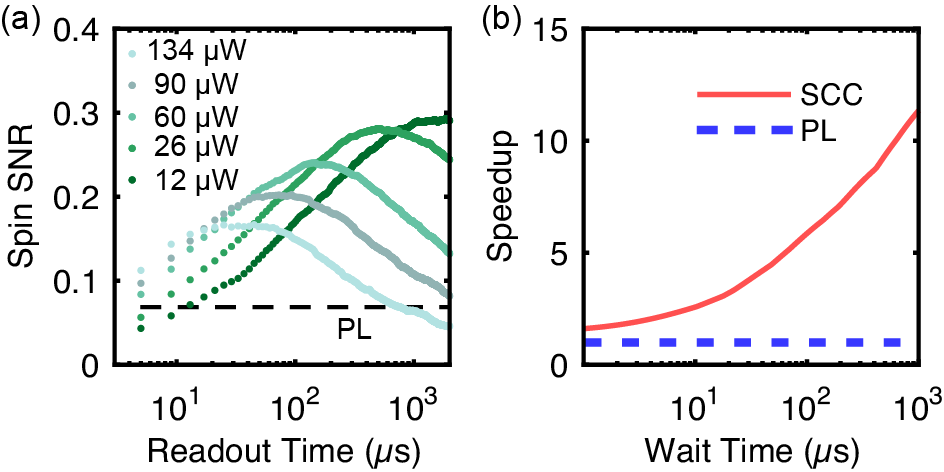}
\caption{\textbf{Agglomerate SCC Calibration and Speedup} (a) Spin SNR calibration of the readout duration and illumination powers. (b) Predicted speed up used to obtain the optimal working parameters for an agglomerate of nanodiamonds,\label{sfig:agglom_scc}}
\end{figure}

SFigure~\ref{sfig:agglom_scc} shows the SCC readout calibration and speedup for an agglomerate of nanodiamonds. The fact that the SNR increase is still present, as well as speedup~$>1$ suggests that SCC is robust across all nanodiamonds studied. This particular calibration was used to generate the qualitative demonstration of SCC spin readout amplification in Fig.1(a) in the main text to increase the observable signal for easier viewing. The agglomerate had a much lower $T_1$ of $150\pm\SI{30}{\micro\second}$ (SFig.~\ref{sfig:t1_mod}). Despite this reduction in wait time, we still achieved a 5-fold speed up for this particular cluster of nanodiamonds.

\bibliography{nd_scc_bibliography.bib}